%% file: main.tex
\documentclass[sigconf,final,screen]{acmart}

\AtBeginDocument{%
  }

\input{preamble/conference_information}
\input{preamble/definitions}
\input{preamble/packages}

\acmSubmissionID{289}

\begin{document}

\title{Generative Retrieval for Book Search}

\input{preamble/authors}

\begin{abstract}
In book search, relevant book information should be returned in response to a query.
Books contain complex, multi-faceted information such as metadata, outlines, and main text, where the outline provides hierarchical information between chapters and sections.
Generative retrieval (GR) is a new retrieval paradigm that consolidates corpus information into a single model to generate identifiers of documents that are relevant to a given query.
How can GR be applied to book search?
Directly applying GR to book search is a challenge due to the unique characteristics of book search: 
\begin{enumerate*}[label=(\roman*)]
    \item The model needs to retain the complex, multi-faceted information of the book, which increases the demand for labeled data.
    \item Splitting book information and treating it as a collection of separate segments for learning might result in a loss of hierarchical information.
\end{enumerate*}

We propose an effective \underline{G}enerative retrieval framework for \underline{B}ook \underline{S}earch (GBS) that features two main components: 
\begin{enumerate*}[label=(\roman*)]
\item data augmentation and
\item outline-oriented book encoding.
\end{enumerate*}
For data augmentation, GBS constructs multiple query-book pairs for training; it constructs multiple book identifiers based on the outline, various forms of book contents, and simulates real book retrieval scenarios with varied pseudo-queries. This includes coverage-promoting book identifier augmentation, allowing the model to learn to index effectively, and diversity-enhanced query augmentation, allowing the model to learn to retrieve effectively.
Outline-oriented book encoding improves length extrapolation through bi-level positional encoding and retentive attention mechanisms to maintain context over long sequences.
Experiments on a proprietary Baidu dataset demonstrate that GBS outperforms strong baselines, achieving a 9.8\% improvement in terms of MRR@20, over the state-of-the-art RIPOR method. 
Experiments on public datasets confirm the robustness and generalizability of GBS, highlighting its potential to enhance book retrieval.
\end{abstract}

\input{preamble/ccs_keywords}


\maketitle

\input{sections/introduction}
\input{sections/methodology}
\input{sections/experimental_settings}
\input{sections/experimental_results}

\input{sections/related_work}
\input{sections/conclusion}

\bibliographystyle{ACM-Reference-Format}
\balance
\bibliography{references}

\input{sections/appendix}

\end{document}

%% file: preamble/conference_information.tex
\copyrightyear{2025}
\acmYear{2025}
\setcopyright{cc}
\setcctype{by}
\acmConference[KDD '25]{Proceedings of the 31st ACM SIGKDD Conference on Knowledge Discovery and Data Mining V.1}{August 3--7, 2025}{Toronto, ON, Canada}
\acmBooktitle{Proceedings of the 31st ACM SIGKDD Conference on Knowledge Discovery and Data Mining V.1 (KDD '25), August 3--7, 2025, Toronto, ON, Canada}
\acmDOI{10.1145/3690624.3709435}
\acmISBN{979-8-4007-1245-6/25/08}

%% file: preamble/definitions.tex
\newcommand{\heading}[1]{\vspace*{1mm}\noindent\textbf{#1.}}

\allowdisplaybreaks

%% file: preamble/packages.tex
\PassOptionsToPackage{numbers,sort&compress}{natbib}

\usepackage[inline]{enumitem} 
\usepackage{acronym}
\usepackage{bbm}
\usepackage{multirow}
\usepackage{multicol}
\usepackage[skip=2pt]{caption}
\usepackage{stfloats}
\usepackage{balance}

%% file: preamble/authors.tex
\author{Yubao Tang}
\affiliation{
 \institution{CAS Key Lab of Network Data Science and Technology, ICT, CAS}
 \institution{University of Chinese Academy of Sciences}
  \institution{University of Amsterdam}
\city{Amsterdam}
  \country{The Netherlands}
}
\email{y.tang3@uva.nl}

\author{Ruqing Zhang}
\authornote{Research conducted when the author was at the University of Amsterdam.}
\affiliation{
 \institution{CAS Key Lab of Network Data Science and Technology, ICT, CAS}
 \institution{University of Chinese Academy of Sciences}
\city{Beijing}
 \country{China}
}
\email{zhangruqing@ict.ac.cn}

\author{Jiafeng Guo}
\authornote{Jiafeng Guo is the corresponding author.}
\affiliation{
 \institution{CAS Key Lab of Network Data Science and Technology, ICT, CAS}
 \institution{University of Chinese Academy of Sciences}
   \city{Beijing}
 \country{China}
}
\email{guojiafeng@ict.ac.cn}

\author{Maarten de Rijke}
\affiliation{
 \institution{University of Amsterdam}
 \city{Amsterdam}
 \country{The Netherlands}
}
\email{M.deRijke@uva.nl}

\author{Shihao Liu}
\affiliation{%
  \institution{Baidu Inc.}
  \city{Beijing}
  \country{China}}
\email{liushihao02@baidu.com}

\author{Shuaiqiang Wang}
\affiliation{%
  \institution{Baidu Inc.}
  \city{Beijing}
  \country{China}}
\email{wangshuaiqiang@baidu.com}

\author{Dawei Yin}
\affiliation{%
  \institution{Baidu Inc.}
  \city{Beijing}
  \country{China}}
\email{yindawei@acm.org}

\author{Xueqi Cheng}
\affiliation{
 \institution{CAS Key Lab of Network Data Science and Technology, ICT, CAS}
 \institution{University of Chinese Academy of Sciences}
   \city{Beijing}
 \country{China}
}
\email{cxq@ict.ac.cn}
\renewcommand{\shortauthors}{Yubao Tang et al.}

%% file: preamble/ccs_keywords.tex
\ccsdesc[500]{Information systems~Retrieval models and ranking}

\keywords{Book retrieval, Generative retrieval, Generative models}

%% file: sections/introduction.tex
\section{Introduction}
\label{section:introduction}

Search engines have become fundamental tools for accessing information in our daily lives. 
As a service offered by generic search engines, book search \cite{xie2020sembrs,wu2008book,ullah2020social} often provides crucial resources for various downstream tasks, including question answering \cite{karpukhin2020dense,nie2020dc} and entity retrieval \cite{genre,chatterjee2021entity}, since books are likely to contain high-quality, authoritative, and comprehensive information.
The increasing demand for book search underscores its importance in meeting user needs effectively \cite{dube2021improving}.

Book search differs significantly from generic web search in two respects \cite{khusro2014issues}: 
\begin{enumerate*}[label=(\roman*)]
    \item Books contain more complex information than web pages, including metadata, outlines, and main text; making effective use of book information requires extensive annotated data. 
    \item Books present extended content with intricate structural relationships. The outline reveals hierarchical relationships between chapters and sections, highlighting a complex interplay of text segments that must be understood for effective retrieval.
\end{enumerate*}

Generative retrieval (GR) \cite{modelBased,DSI} is an emerging retrieval paradigm that integrates all corpus information into a consolidated model that is capable of directly generating relevant document identifiers (docids) for queries. To achieve this, GR involves two fundamental learning tasks: 
\begin{enumerate*}[label=(\roman*)]
    \item indexing, which includes learning associations between books and docids,  and 
    \item retrieval, which maps queries to their corresponding docids. 
\end{enumerate*}
GR has garnered increasing attention for its robust performance in document, passage, and entity retrieval tasks \cite{genre,chen2022corpusbrain,zeng2023scalable,zeng2024planning,tang-2023-recent,tang2023semanticenhancedsedsi}.

Directly applying GR to book search is a challenge due to the unique characteristics of books, as described in the abstract. Simply splitting book content into multiple independent segments for GR learning may result in a loss of hierarchical relationships inherent in book structures, thereby affecting retrieval quality. The use of GR for book search remains a complex and unresolved problem.

Our objective is to develop an effective GR framework for book search, called GBS, that can accurately return relevant book identifiers for given queries. To accomplish this, we need to address two key challenges. 

First, \textit{how to achieve data augmentation under the GR framework?} Considering the two learning tasks of GR (indexing and retrieval), we propose a data augmentation method that constructs various training data pairs for indexing and retrieval. 
This method includes: 
\begin{enumerate*}[label=(\roman*)]
    \item Coverage-promoting book identifier augmentation for indexing, which aims to improve the coverage of book identifiers by using the outline information to design hierarchical book identifiers and various forms of book content, namely book-, chapter- and section-level identifiers. We further train a model to learn the mapping between these diverse content forms and their corresponding hierarchical book identifiers, ensuring that the model comprehensively understands the structure and content of the books.
 
    \item Diversity-enhanced query augmentation for retrieval, which considers queries of varying difficulty levels, from those that can be answered by a single chapter to those requiring content across multiple chapters. To generate high-quality queries that meet these criteria, we design different types of prompts using large language models (LLMs) with strong text generation capabilities. These prompts help in creating a rich and varied set of pseudo-queries, enhancing the training data's diversity and quality. These queries are then paired with book-level identifiers used in learning the retrieval.
    
\end{enumerate*}

Second, \textit{how to improve length extrapolation for long book input?} 
If the book content is input sequentially, it is likely to lose the hierarchical relationships between chapters and sections provided by the outline. Additionally, there is a limitation on the length of input that the model is able to handle. 
To address these issues, it is crucial to incorporate outline information. Not only does the outline provide hierarchical context, but it also structures the content, which can help manage the length of information input.
Therefore, we propose an outline-oriented book encoding that hierarchically distinguishes the chapter and section content of a book based on its outline structure. 
This approach includes: 
\begin{enumerate*}[label=(\roman*)]
    \item Outline-oriented bi-level positional encoding, which applies hierarchical positional encodings to chapter-level and section-level texts based on the book's outline. This method better captures the relationships between different chapters and sections, reflecting their structural hierarchy. 
    
    \item Outline-oriented retentive attention, which introduces an additional memory module into the traditional multi-head attention mechanism. This module stores important history information from the input text, helping the attention mechanism to more effectively filter and integrate critical information from long texts. This enhancement allows the model to maintain context over longer sequences, which is essential for accurately processing the extensive content within books.
\end{enumerate*}

\begin{figure}[t]
     \centering
     \includegraphics[width=0.4\textwidth]{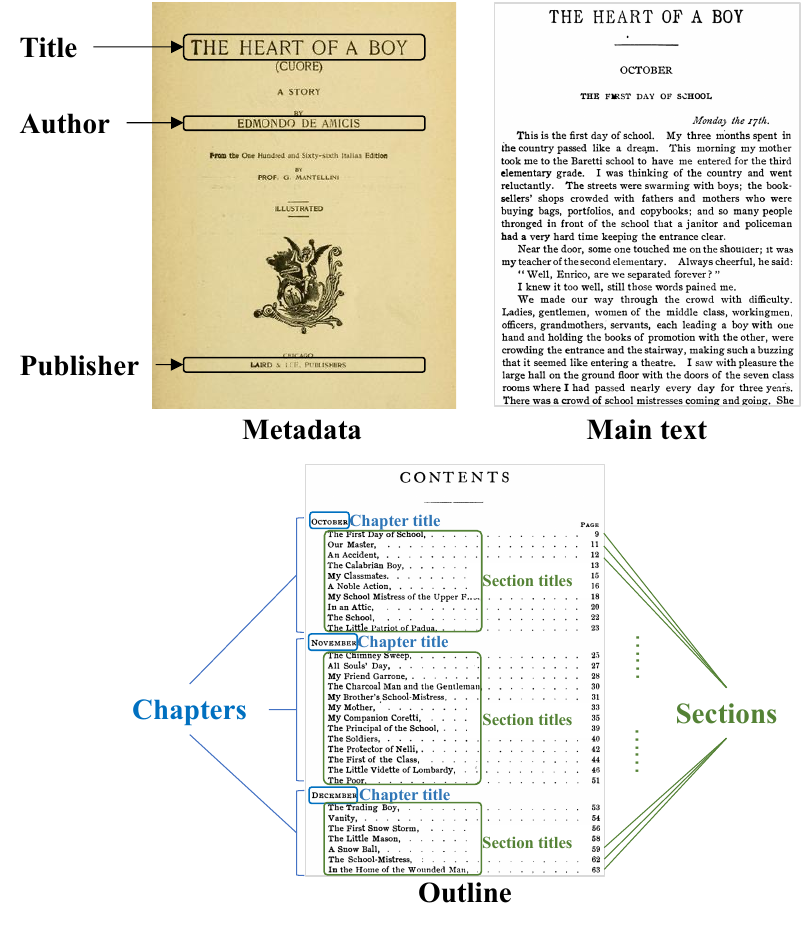}
     \caption{Books mainly consist of three types of information: (1) metadata, which includes details like the title, author, and publisher; (2) the main text, which constitutes the core content of the book; and (3) the outline, which shows the hierarchical structure and relationships between the chapters and sections.}
     \Description{Book resource}
     \label{fig:book-resource}
\end{figure}

\begin{figure*}[t]
    \centering
    \includegraphics[width=0.9\textwidth]{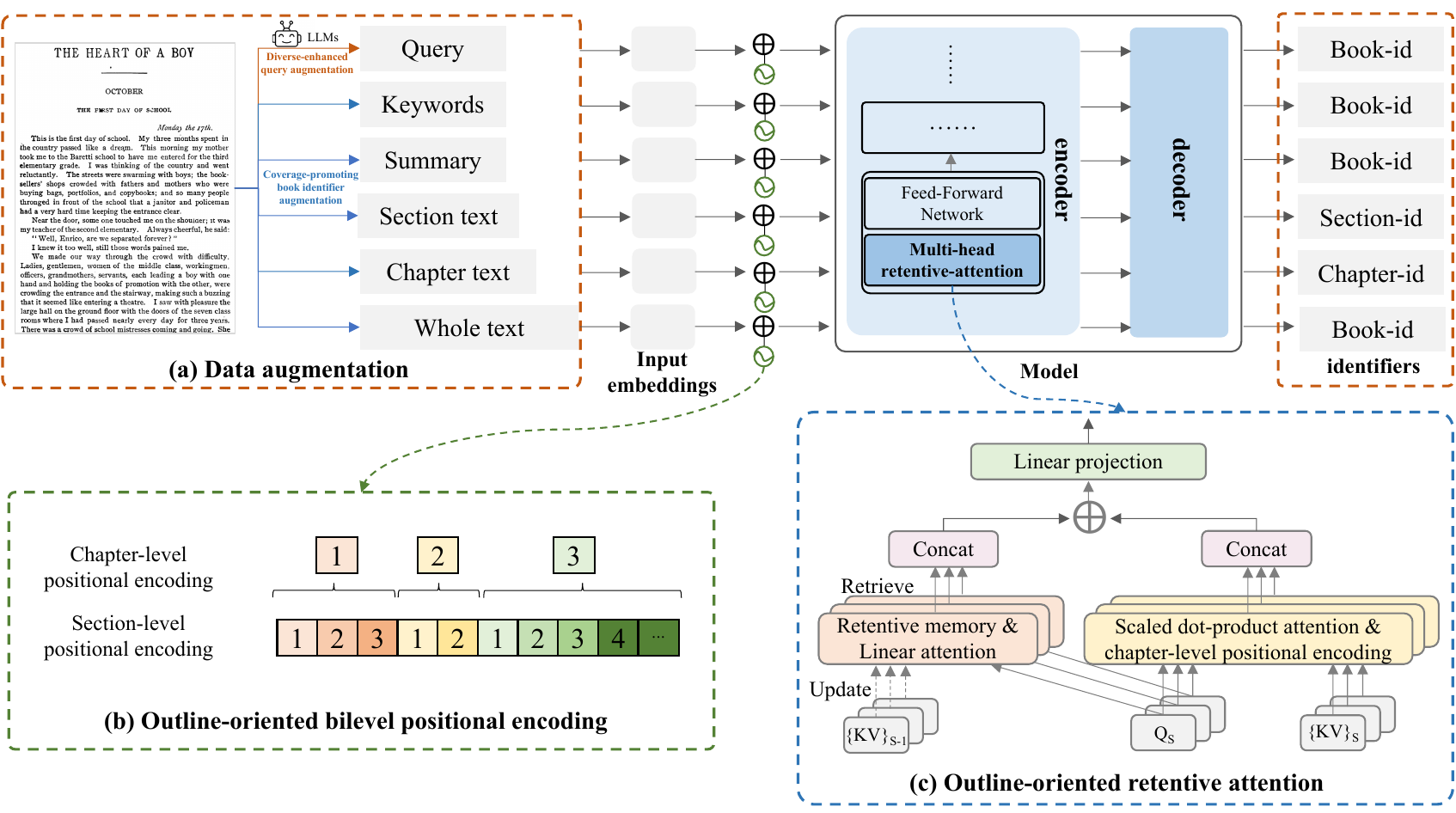}
    \caption{Based on an encoder-decoder architecture, GBS comprises two components: (1) Data augmentation (orange dashed rectangles), which includes coverage-promoting book identifier augmentation for indexing and diverse-enhanced query augmentation for retrieval, generating multiple data pairs. (2) Outline-oriented book encoding, which includes outline-oriented bi-level positional encoding (green dashed rectangles) and outline-oriented retentive attention (blue dashed rectangles), to encode the long book contents based on hierarchical information. (The figure should be viewed in color.)}
    \Description{Overview}
    \label{fig:overview}
\end{figure*}

Experiments on a proprietary Baidu dataset demonstrate that our proposed method GBR significantly outperforms the state-of-the-art GR baseline, RIPOR, with a 9.8\% improvement in terms of MRR@20. Additionally, we validate the effectiveness of our method on public datasets, WhatsThatBook \cite{lin2023decomposing}, confirming its robustness and generalizability. These results underscore the potential of our GBS to revolutionize book search by addressing its unique challenges and using the full scope of available information.

%% file: sections/methodology.tex
\section{Methodology}

\subsection{Problem statement}
Books are complex and rich information sources,
comprising diverse elements. We denote the book set as $\mathcal{B}=\{b_1, \ldots, b_{|\mathcal{B}|}\}$,  where $b_i$ is the $i$-th book in $\mathcal{B}$.
As shown in Figure \ref{fig:book-resource}, a book includes:
\begin{itemize}[leftmargin=*,nosep]
    \item Metadata: This consists of essential information about the book, such as title, author, publisher, and other bibliographic details. Metadata provides a quick reference to identify books.
  
    \item Outline: This represents structural information about the book, detailing the organization of the contents into chapters and sections. It offers a hierarchical view of the book's contents, helping to understand the flow and structure of information.

    \item Main text: This contains the detailed text, 
    across various chapters and sections. It is where the substantial information resides, providing the comprehensive material covered by the book.

\end{itemize}

\noindent%
We aim to develop an autoregressive GR model $M$ that returns relevant book identifiers given a query $q_i$ in the query set $\mathcal{Q} = \{q_1, \ldots, q_{|\mathcal{Q}}|\}$. 
$M$ needs to accurately understand and interpret the query, using the metadata, main text of books and outline, to identify and return the most relevant book identifiers.

\subsection{Model architecture}\label{sec:model-architecture}
Similar to previous generative retrieval (GR) research \cite{DSI, NCI, zeng2023scalable,genre}, we employ a transformer-based model comprising two main components:
\begin{enumerate*}[label=(\roman*)]
    \item An encoder: A bidirectional encoder equipped with our specially designed multi-head retentive-attention mechanism to encode book contents or pseudo-queries. We provide a detailed explanation of this multi-head retentive-attention mechanism in Section \ref{sec:context-processing}.   
    \item An identifier decoder: This component operates through a sequential generation process to produce book identifiers.
\end{enumerate*}

\subsection{Data augmentation}\label{sec:data-augmentation}
The core idea is to simulate diverse queries in real book retrieval scenarios and to ensure that book identifiers comprehensively represent book information. This allows us to construct multiple training data pairs, thereby fully using book information. The method includes coverage-promoting book identifier augmentation for  indexing and diversity-enhanced query augmentation for retrieval.

\subsubsection{Coverage-promoting book identifier augmentation for indexing}
Given the extensive length of book content, we design multiple forms of book information as input and hierarchical identifiers as output to help the GR model better remember the mapping between book content and identifiers.

We use five types of book content based on the degree of information compression:
\begin{itemize}[leftmargin=*,nosep]
    \item Keywords: A set of words containing key information, directly helping the model to learn the most critical information of a book. These words are extracted from the leading chapter's text using an existing keyword extractor, considering that the leading part of a book usually contains important information.
    \item Summary: Similar to keywords but providing a more coherent and information-rich summary. This is generated by a summarizer from the leading chapter's text.
    \item Section text: The content of a section is relatively complete and contains more details. To minimize information loss, the model needs to learn the fine-grained section text.
    \item Chapter text: Compared to section text, chapters include multiple sections, capturing the relationships between sections and more comprehensive details.
    \item Whole text: The entire text of the book, providing higher-level connections between chapters.
\end{itemize}

\noindent%
   For book identifiers $\mathcal{U}_i$ for book $b_i$, we design three hierarchical levels based on the book's outline structure:
   \begin{itemize}[leftmargin=*,nosep]
    \item Book-level identifier (book-id): It is composed of the book title, author, and publisher, separated by ``\#''. 
    \item Chapter-level identifier (chapter-id): Built on the book-id, we add the chapter title and a chapter-level semantic structured number, separated by ``\#'', to form the chapter-id. Inspired by \cite{DSI}, this chapter-level semantic structured number is obtained by applying a hierarchical K-means clustering algorithm to the text of all chapters in $b_i$. 
    \item Section-level identifier (section-id): Built on the chapter-id of the chapter to which the section belongs, we add the section title and a section-level semantic structured number, separated by ``\#'', to form the section-id. Similarly, this section-level semantic structured number is obtained by applying a hierarchical K-means clustering algorithm to the text of all sections within the current chapter.
   \end{itemize}

\noindent%
Some books might contain finer-grained subsections, but compared to sections, the information in subsections is too fragmented, so we do not consider them here. Additionally, some books may only include the highest-level chapters, in which case we only consider chapter-level identifiers.

Using the methods described above, we can construct multiple data pairs of book content and identifier for learning the indexing task, where the book content is used as the input, and the identifier is used as the output.
Specifically, the whole text, keywords, and summaries correspond to book-id; section text corresponds to section-id; and chapter text corresponds to chapter-id. 
The training methodology is described in Section \ref{sec:training}.

\subsubsection{Diversity-enhanced query augmentation for retrieval}
According to actual retrieval needs, we classify queries into two categories:
\begin{enumerate*}[label=(\roman*)]
    \item Single-chapter answerable queries, which focus only on the details of a single chapter, and 
    \item Multiple-chapter answerable queries, which require the context from multiple chapters to answer comprehensively.
\end{enumerate*}
To generate high-quality queries that meet these criteria, we use LLMs with strong text generation capabilities, designing specific prompts to guide the LLMs in the generation process. The two types of prompts are as follows:
\begin{itemize}[leftmargin=*,nosep]
\item The prompt for single-chapter answerable queries: ``\texttt{Given the following chapter from a book, generate \{X\} pseudo queries that can be answered using the information contained within this single chapter. The queries should focus on key themes, events, characters, and any specific details provided in the chapter. A single chapter content: \{chapter texts\}.}''

\item The prompt for multiple-chapter answerable queries: ``\texttt{Given the following chapters from a book, where they are separated by a token ``\#'', generate \{X\} complex pseudo queries that require synthesizing information from multiple chapters to answer. Each query should be clear, specific, and necessitate the integration of information across different chapters.
Multiple chapter contents: \{chapter texts\}.}''
\end{itemize}
\texttt{\{X\}} and \texttt{\{chapter texts\}} denote the number of pseudo queries and chapter texts, respectively.
Based on this strategy, we can construct multiple data pairs of pseudo-query and relevant book identifier for the retrieval task.

\subsection{Outline-oriented book encoding}\label{sec:context-processing}
The core idea is to use the structural information provided by the book outline to encode the long book content, especially the whole text, as a complete unit. We enhance the positional encoding and the attention mechanism in the transformer encoder by designing outline-oriented bi-level positional encoding and outline-oriented retentive attention.

\subsubsection{Outline-oriented bi-level positional encoding}
The key idea is to encode the whole text input into the model according to the outline, highlighting the relationships between different chapters and sections, through bi-level positional encoding, including section- and chapter-level positional encoding.

\heading{Section-level positional encoding}
In an input book's complete text sequence $L=[w_1, \ldots, w_{|L|}]$, each section is viewed as an independent segment unit. The section-level positional encoding is used to pinpoint the position of each token within the section, facilitating the capture of semantic information. Formally, within each section $L_l=[w_{a_l}, w_{a_l+1}, \ldots, w_{b_l}]$ where $l \in |L|$ and $a_l$ and $b_l$ are the starting and ending indices, we encode the (local) position $j$ for token $w_{a_l+j}$, with $1 \leq j \leq b_l - a_l + 1$. Typically, the number of tokens within a section is limited, making the original absolute positional encoding sufficient \cite{vaswani2017attention}.
Each token $w_{a_l+j}$ in $L_l$ is assigned a real-valued embedding $e_j$, which is then added to the input token embedding. This embedding $e_j$ is consistent across tokens at the same local position $j$ in different sections $L_l$. The section-level positional encoding is combined with the input embedding, helping to maintain the structural and contextual integrity within the book.

\heading{Chapter-level positional encoding}
Although the section-level positional encoding pinpoints token locations within individual sections, it does not differentiate locations across different sections, failing to capture inter-sectional contextual relationships. To address this, we further introduce chapter-level positional encoding, which specifies the section each token belongs to, enhancing the handling of longer sequences that are not present during training. This chapter-level encoding uses relative positional encodings based on the distance between section indexes  \cite{su2024roformer, shaw2018self}.
We adopt rotary position encoding \cite{su2024roformer} as our chapter-level positional encoding. For a pair of tokens $(w_{l_1}, w_{l_2})$ located in the $n$-th and $m$-th sections respectively, we assign two rotation matrices $R_{f,n}$ and $R_{f,m}$, where $f$ represents the predefined parameters of the rotation matrix \cite{su2024roformer}. Given an attention query-key pair $\hat{q}_{l_1}$ and $k_{l_2}$ in $\mathbbm{R}^d$, the attention score is calculated as 
$(\hat{q}_{l_1}R_{f,n}(k_{l_2}R_{f,m})^T)\cdot (\sqrt{d})^{-1} = (\hat{q}_{l_1}R_{f,n-m}k_{l_2}^T)\cdot (\sqrt{d})^{-1}$. Here, $\hat{q}$ is the attention query, distinct from the constructed pseudo-query used as the model's input, and $d$ is the dimension and scaling factor of $k$. This chapter-level positional encoding is integrated into the original standard multi-head attention (MHA) module in the transformer encoder \cite{vaswani2017attention}.

\subsubsection{Outline-oriented retentive attention}
We introduce an additional retentive memory module into the encoder of the transformer-based backbone \cite{vaswani2017attention,raffel2020exploringt5}. This module stores crucial information from the input text, which is then aggregated with the original standard MHA. This helps the attention mechanism more effectively filter and integrate important information from long texts.

\heading{Standard multi-head attention}
For an individual head in the standard MHA, the attention context $C \in \mathbbm{R}^{N \times d}$ is calculated by scaled dot-product attention, which is derived from a sequence of input texts $L \in \mathbbm{R}^{N \times d}$ using the formula $C = \mathrm{softmax} ( (\hat{Q}K^T)\cdot (\sqrt{d})^{-1} ) V$. In this context, $\hat{Q} = L\xi_{Q}$, $K = L\xi_{K}$, and $V = L\xi_{V}$, where $\xi_{Q}$, $\xi_{K}$, and $\xi_{V}$ are trainable projection matrices.
In MHA, we generate $H$ attention context vectors for each element in the sequence simultaneously, concatenate these vectors along the second dimension, and project the concatenated vector back to the model space to produce the final attention output.

\heading{Retentive memory}
In retentive attention, rather than generating new memory entries, we reuse the query, key, and value states ($\hat{Q}$, $K$, and $V$) obtained from the dot-product attention process. This reuse of states between dot-product attention and retentive memory facilitates efficient adaptation to long contexts and enhances both training and inference speed. The aim is to store the key-value pairs in the retentive memory and retrieve them using query vectors, following \cite{munkhdalai2019metalearned}. The retentive memory is parameterized with an associative matrix \cite{schlag2020learning}, allowing the update and retrieval process to be framed as a linear attention mechanism \cite{shen2021efficient}. This method benefits from stable training techniques applied in similar methods. We specifically use the update and retrieval approach from \cite{katharopoulos2020transformers} due to its simplicity and effectiveness.

\textbf{Retentive memory retrieval. }
The new content $C_\mathit{new}$ from the retentive memory $MM_{s-1}$ is computed using $\hat{Q}$ as 
$ C_\mathit{new} = (\sigma(\hat{Q}) \cdot MM_{s-1})\cdot (\sigma(\hat{Q}) \cdot z_{s-1})^{-1}$.
Here, $\sigma$ denotes an element-wise ELU + 1 activation function \cite{clevert2015fast}, and $z_{s-1}$ represents a normalization term that is the sum over all keys, following \cite{katharopoulos2020transformers}.

\textbf{Retentive memory update.} 
After retrieval, we update the memory and normalization term with the new key-value entries. The updated states are calculated as follows:
\begin{align}
    MM_s \leftarrow MM_{s-1} + \sigma(K)^T V , \\
z_s \leftarrow z_{s-1} + \sum_{t=1}^N \sigma(K_t).
\end{align}
The newly computed memory states, $MM_s$ and $z_s$, are then passed to the subsequent segment unit $s+1$, establishing a recurrence within each attention layer.

\textbf{History context injection.} 
We combine the local attention state $C$ and the retrieved memory content $C_{new}$ using a learned gating scalar $\alpha$ as follows: 
\begin{equation}
    C_{total} = \text{sigmoid}(\alpha) \odot C_{new} + (1 - \text{sigmoid}(\alpha)) \odot C.
\end{equation}
In multi-head retentive attention, we compute $H$ context states in parallel. These states are then concatenated and projected to produce the final attention output:
$A = [C^1_\mathit{total}; \ldots; C^H_\mathit{total}] \xi_A$, where $\xi_A$ represents the trainable weights.

\subsection{Training}\label{sec:training}
Following \cite{DSI,NCI}, for the constructed training data pairs in Section \ref{sec:data-augmentation}, we adopt a maximum likelihood estimation (MLE) \cite{xiongdcn+mle} to learn the indexing and retrieval tasks.
For the indexing task, given various forms of book as input, the model maximizes the likelihood of the corresponding identifiers as output. Keywords, summaries, and the whole text correspond to book-id; section text corresponds to the section-id; and chapter text corresponds to the chapter-id. This task can be formalized as:
\begin{equation}
    \mathcal{L}_{ind}(\mathcal{B},\mathcal{Y};\theta) = -\sum^{|B|}_{b_i \in B, y_{i,j} \in \mathcal{Y}_i, o_{i,j} \in \mathcal{O}_{i}} \log P(o_{i,j}|y_{i,j}),
\end{equation}
where $y_{i,j} \in \mathcal{Y}_i$ can be any form of input for book $b_i$, and $o_{i,j} \in \mathcal{O}_i$ is the corresponding identifier for $y_{i,j}$.

For the retrieval task, given pseudo-queries as input, the model maximizes the likelihood of the corresponding book identifiers as output, formalized as:
\begin{equation}
    \mathcal{L}_{rel}(\mathcal{B},\mathcal{Q};\theta) = -\sum^{|B|\times |\mathcal{Q}|}_{b_i \in B, q_{i,j} \in \mathcal{Q}} \log P(u_{i}^t|q_{i,j}),
\end{equation}
where $\theta$ represents the model parameters, and $u_{i}^t$ is the relevant book-id corresponding to book $b_i$, of query $q_{i,j}$.

We employ a multi-task learning approach to train these two tasks together, following \cite{DSI,zhuang2022bridgingdsiqg}. The overall optimization objective can be formalized as:
\begin{equation}
    \mathcal{L}(\mathcal{B},\mathcal{Y}, \mathcal{Q};\theta) =\mathcal{L}_{ind}(\mathcal{B},\mathcal{Y};\theta) + \mathcal{L}_{rel}(\mathcal{B},\mathcal{Q};\theta).
\end{equation}

\subsection{Inference}\label{sec:inference}
After training the model, we proceed with inference when a query is input. Following \cite{genre}, to ensure that the generated identifiers are valid, we constrain the model's generation using prefix trees. To integrate different levels of identifiers, we consider book-level and chapter-level identifiers. Given the excessive number of section-level identifiers for a single book, which could confuse the model during inference, we temporarily exclude this level. We 
\begin{enumerate*}[label=(\roman*)]
    \item first use the identifiers to construct prefix trees, then 
    \item perform decoding using parallel or serial methods, and aggregate the inference results from both levels to obtain the relevance score between the book and the input query.
\end{enumerate*}

\heading{Prefix tree construction}  
For the prefix tree, nodes are annotated with tokens from the predefined candidate set. Each node's children represent all allowed continuations from the prefix defined by traversing the tree from the root to it \cite{genre}. We construct three types of tree:
\begin{itemize}[leftmargin=*,nosep]
    \item Book-level prefix tree: It includes all book-ids in $B$.
    \item Individual book prefix tree: We construct an individual tree for each book's chapter-ids.
    \item Chapter-level prefix tree: We construct a tree from all books' chapter-ids together.
\end{itemize}

\heading{Parallel decoding and aggregation}
First, we decode using the book-level prefix tree and the chapter-level prefix tree separately to obtain the book-level identifier list (book-id list) and the chapter-level identifier list (chapter-id list). Each book's book-level relevance score is its corresponding probability likelihood value, denoted as $s_{b}(q,b_i)$. Each book's chapter-level relevance score is the total number of its chapter-ids covered in the generated chapter-id list, denoted as $s_{c}(q,b_i)$.

Second, following \cite{li2023multiview}, the chapter-level relevance score acts as a weight for the book-level relevance score. We aggregate the overall relevance score between a query $q$ and a book $b_i$ as the product of the two scores, formalized as 
$score(q,b_i)=s_{b}(q,b_i) \times s_{c}(q,b_i)$.

\heading{Serial decoding and aggregation}
First, we perform constrained decoding using the book-level prefix tree to obtain the relevant book-id list. Each book's book-level relevance score is as in parallel decoding, denoted as $s_{b}(q,b_i)$.

Next, within the scope of the relevant book-id list, we perform further fine-grained inference for each book using its individual book prefix tree to obtain the relevant chapter-id list. Each book's chapter-level relevance score is the sum of the likelihood values of all its chapter-ids in the list, denoted as $s_{c}(q,b_i)$.

Finally, we aggregate the overall relevance score between $q$ and $b_i$ as the weighted sum of the two scores, formalized as 
$score(q,b_i)=\beta s_{b}(q,b_i) + \gamma s_{c}(q,b_i)$, where $\beta$ and $\gamma$ are the weights. 



\subsection{GBS}
GBS is defined as follows: first, we construct data pairs for books using the data augmentation described in Section \ref{sec:data-augmentation}. 
For each pair, we encode the input based on the model structure described in Section \ref{sec:model-architecture} using the outline-oriented book encoding described in Section \ref{sec:context-processing}, and then predict the identifier through the decoder. This process is optimized using the training objective described in Section \ref{sec:training}. After training, the model is used for inference with any decoding method (parallel or serial) described in Section \ref{sec:inference}.
We consider two variants of GBS, GBS$^P$ (which uses parallel decoding) and GBS$^S$ (which uses serial decoding).

%% file: sections/experimental_settings.tex
\section{Experimental Settings}

\heading{Datasets}
We use both a proprietary dataset and a public dataset for our experiments.
\begin{enumerate*}[label=(\roman*)]
    \item Baidu book search (BBS) dataset: This dataset is from Baidu's real-world scenario and includes a library of books with metadata, outlines, and main text. 
    The dataset contains both Chinese and English books. 
    We sampled three datasets of different scales, namely 10K, 20K, and 40K. On average, each book contains about 225K words. We construct pseudo-queries for each book for training and evaluation using the method described in Section \ref{sec:data-augmentation}. Specifically, we generated five single-chapter answerable queries and five multiple-chapter answerable queries for each book, i.e., $X=5$. 

    \item WhatsThatBook \cite{lin2023decomposing}: This dataset consists of tip-of-the-tongue queries for book searches, collected from user interactions on the GoodReads\footnote{\url{https://www.goodreads.com/}} community forum. This dataset contains only English books and queries.
    The queries include the forum discussions, and the documents are books with their corresponding metadata. On average, each book contains about 131K words.
    Note that the original training queries in this dataset total 11.6K, with an average of one annotated query per book. Additionally, we generate 4 pseudo-queries of each of the two types for every book.
\end{enumerate*}

The dataset statistics are provided in Table \ref{tab:data-statistics}.

\begin{table}[t]
    \caption{Statistics of datasets. \#Book denotes the number of books. \#Train denotes the number of the queries in the training set. \#Test denotes the number of queries for testing.}
    \label{tab:data-statistics}
    \centering
    \renewcommand{\arraystretch}{0.9}
    \setlength\tabcolsep{10pt}
    \begin{tabular}{lrrr}
         \toprule
        \textbf{Dataset} & \textbf{\#Book}  & \textbf{\#Train} & \textbf{\#Test} \\
        \midrule
       \textbf{BBS 10K} &10K &1M & 1K \\
       \textbf{BBS 20K} & 20K&2M & 1K\\
       \textbf{BBS 40K} &40K &4M &1.5K \\
       \textbf{WhatsThatBook} &14K &1.5M & 1.45K \\
        \bottomrule
    \end{tabular}
    \end{table}

\heading{Evaluation metrics}
In line with the GR work by \cite{DSI, li2023multiview,tang2023semanticenhancedsedsi}, we adopt hit ratio (Hits$@K$) with $K=\{10\}$ and mean reciprocal rank (MRR$@K$) with $K=\{20\}$ as our evaluation metrics.

\heading{Baselines}
Following existing GR research \cite{DSI,zeng2023scalable,wang2023novo}, we consider three types of baselines as follows:
\begin{enumerate*}[label=(\roman*)]
    \item \emph{Sparse retrieval baselines}: BM25 \cite{bm25}, and DocT5Query \cite{doct5query}.

    \item \emph{Dense retrieval baselines}: RepBERT \cite{zhan2020repbert}, and DPR \cite{karpukhin2020dense}. 

    \item \emph{GR baselines}: DSI \cite{DSI}, GENRE \cite{genre}, SEAL \cite{seal}, DSI-QG \cite{zhuang2022bridgingdsiqg}, NCI \cite{NCI},  Corpusbrain \cite{chen2022corpusbrain}, Ultron \cite{zhou2022ultron}, GenRet \cite{sun-2023-learning-arxiv}, NOVO \cite{wang2023novo}, ASI \cite{yang2023auto} and RIPOR \cite{zeng2023scalable}. 
\end{enumerate*}
For the dense retrieval and GR baselines, we split the text of a book into multiple segments, and then form multiple data pairs of the book segment and query for training.

Additionally, for the GR baselines, we also construct multiple pairs of the book segment the corresponding identifier for the indexing task.
All GR baselines are optimized with an encoder-decoder architecture using MLE. 
For more details on our baselines, please refer to Appendix~\ref{appendix-baselines}.

\subsection{Implementation details}

\heading{Backbone} 
For the Baidu book retrieval dataset, which contains both Chinese and English books, we adopt Mengzi-T5-base \cite{zhang2021mengzi}, a language model pretrained on both Chinese and English corpora, as the backbone. For the WhatsThatBook dataset, since it is entirely in English, we used the T5-base model \cite{raffel2020exploringt5}, which is widely used in the GR research \cite{DSI,NCI,zeng2023scalable}, as the backbone.
For both models, the hidden size is 768, the feed-forward layer size is 3072, the number of self-attention heads is 12, and the number of transformer layers is 12. Decoder-only architectures, such as the GPT series models \cite{2022Traininggpt}, will be explored in future research.

\heading{Hyperparameters}
Regarding chapter- and section-level semantic structured numbers, following 
\cite{DSI}, we encode the text using a small 8-layer BERT model and set the number of clusters to 10, with a maximum threshold of 100 for each layer. And for serial decoding, we set $\beta$ as 1 and $\gamma$ as 0.5.

\heading{Training and inference}
GBS is implemented with PyTorch 1.9.0 and HuggingFace transformers 4.16.2; we re-implement DSI, and use open-source code for other baselines. 
For the model that extracts book keywords and summaries, we use the TextRank model, implemented via the summa API \cite{DBLP:journals/corr/BarriosLAW16}. The whole text in the (whole text, book-id) pairs used for training consists of the first 100 chapters of the book.
During inference
We employ the Adam optimizer with a linear warm-up over the initial 10\% of steps. The learning rate is set to 5e-5, with a label smoothing of 0.1, weight decay of 0.01, a maximum of 5M training steps, and a batch size of 128. 
We set the input length to 128K, truncating any portion of the book that exceeds this limit.
Our model is trained on eight NVIDIA Tesla A100 80GB GPUs. 
, we use constrained beam search with 20 beams to decode the identifiers.


%% file: sections/experimental_results.tex
\section{Experimental Results}
This section presents the experimental findings. 

\begin{table*}[t]
    \centering
    \setlength{\tabcolsep}{5pt}
    \caption{Retrieval performance on BBS and WhatsThatBook. The best results are shown in \textbf{bold}.  $\ast$ indicates statistically significant improvements over the best performing GR baseline RIPOR  ($p \leq 0.05$).}
    \label{tab:main-results}
    \renewcommand{\arraystretch}{0.95}
    \begin{tabular}{l l ccccc ccccc}
        \toprule
        & \multirow{2}{*}{\textbf{Method}} &  
        \multicolumn{2}{c}{\textbf{BBS 10K}} & 
        \multicolumn{2}{c}{\textbf{BBS 20K}}& 
        \multicolumn{2}{c}{\textbf{BBS 40K}}& 
        \multicolumn{2}{c}{\textbf{WhatsThatBook}}
        \\
        \cmidrule(r){3-4}
        \cmidrule(r){5-6}
        \cmidrule(r){7-8}
        \cmidrule(r){9-10}

        & & \textbf{Hits@10} & \textbf{MRR@20} & \textbf{Hits@10} & \textbf{MRR@20} & \textbf{Hits@10} & \textbf{MRR@20} & \textbf{Hits@10} & \textbf{MRR@20} \\
        \midrule 
        \multirow{2}{*}{\rotatebox[origin=c]{90}{\small Sparse}}
        & BM25 &41.8 & 30.5 & 40.6 & 30.1 & 40.1 & 29.8 & 45.3 & 41.6\\
        & DocT5query & 46.6 & 39.3 & 42.5 & 35.4 & 37.5 & 30.7 & 48.2 & 42.8\\
        \midrule
        \multirow{2}{*}{\rotatebox[origin=c]{90}{\small Dense}}
         & RepBERT & 53.1 & 46.4 & 48.3 & 41.9 & 45.6 & 37.3 & 56.8 & 48.1\\
         & DPR &  51.3 & 43.6 & 46.4 & 40.3 & 42.1 & 35.9 & 54.2 & 45.7\\
        \midrule
        \multirow{11}{*}{\rotatebox[origin=c]{90}{\small Generative}}
         & DSI &20.7 & 13.4 & 18.5 & 11.6 & 13.5 & 7.2 & 22.6 & 15.4 \\
         & GENRE &26.5 & 22.1 & 24.5 & 19.3 & 19.3 & 15.6 & 29.1 & 24.7\\
         & SEAL   & 27.8 & 23.6 & 25.7 & 20.6 & 20.5 & 16.2 & 30.5 & 24.8\\
         & DSI-QG &40.3 & 35.7 & 36.8 & 28.9 & 32.1 & 23.7 & 44.6 & 25.9 \\
         & NCI  & 42.8 & 36.8 & 37.2 & 29.4 & 32.8 & 24.2 & 45.2 & 39.4\\
         & Corpusbrain & 48.9 & 43.1 & 42.4 & 37.5 & 36.3 & 31.5 & 52.3 & 45.7\\
         & Ultron &48.6 & 44.5 & 41.3 & 36.3 & 35.7 & 30.1 & 51.9 & 45.2\\
         & GenRet&51.3 & 46.6 & 47.2 & 42.1 & 41.4 & 37.8 &55.8 &	49.3 \\
         & NOVO&52.7 & 47.3 & 47.6 & 42.8 & 41.8 & 38.4 & 56.5 &	49.7 \\
         & ASI &58.4 & 53.6 & 53.5 & 44.6 & 46.6 & 40.1 &56.2 &	49.5 \\
         & RIPOR& 62.5 & 52.8 & 56.8 & 46.2 & 52.9 & 42.7 & 66.7 & 55.4 \\
         \midrule
         \multirow{2}{*}{\rotatebox[origin=c]{90}{\small Ours}}
        & GBS$^S$ &66.4 & 54.1 & 61.3 & 49.5 & 56.3 & 46.5 &  70.4 & 58.1\\
         & GBS$^P$ &\textbf{66.7}\rlap{$^{\dagger}$} & \textbf{54.4}\rlap{$^{\dagger}$} & \textbf{61.6}\rlap{$^{\dagger}$} & \textbf{49.8}\rlap{$^{\dagger}$} & \textbf{56.7}\rlap{$^{\dagger}$} & \textbf{46.9}\rlap{$^{\dagger}$} & \textbf{70.7}\rlap{$^{\dagger}$} & \textbf{58.6}\rlap{$^{\dagger}$} \\
        \bottomrule
    \end{tabular}
\end{table*}

\subsection{Main results}
A comparison between the proposed GBS and baselines on the BBS and WhatsThatBook datasets is shown in Table \ref{tab:main-results}.

\heading{Performance of sparse retrieval and dense retrieval baselines}
\begin{enumerate*}[label=(\roman*)]
    \item BM25 shows stable performance across the three scales of the BBS dataset. However, performance slightly declines as the number of books increases. This decline might be due to the fact that dividing a book into multiple segments for indexing increases the number of thematically similar fragments, making retrieval more challenging.

    \item DocT5query performs better overall than BM25. This may be because the generated pseudo-queries are based on the leading chapters of the books, potentially including more key information from the books.

    \item The two dense retrieval baselines, RepBERT and DPR, outperform the sparse retrieval baselines. This is likely because dense embeddings capture more semantic information.
\end{enumerate*}

\heading{Performance of GR baselines}
\begin{enumerate*}[label=(\roman*)]
    \item DSI performs worse than sparse retrieval baselines, possibly because the information in books is too rich. Using only a semantically structured number as identifiers may lose too much information. Although GENRE improves over DSI, it still performs relatively poorly, likely because using book titles as identifiers, while containing more information than a semantically structured number, still represents only one identifier, leading to information loss.

    \item DSI-QG and NCI show significant improvements over the aforementioned GR baselines. These improvements may be due to their use of additional generated pseudo-queries during training.

    \item Corpusbrain and Ultron achieve further improvements due to pre-training of the models.

    \item GenRet, NOVO, and ASI perform similarly to dense retrieval baselines, with ASI even achieving better results. This is likely because these three baselines specifically learn suitable identifiers for the retrieval.

    \item RIPOR outperforms other baselines, possibly because its multi-stage learning and negative sampling strategies enhance the model’s understanding of the book corpus and relevance.
\end{enumerate*}

\heading{Performance of GBS}
\begin{enumerate*}[label=(\roman*)]
    \item  Both variants, GBS$^S$ and GBS$^P$, achieve better results than the baselines. Specifically, GBS$^P$ outperforms the best baseline, RIPOR, by 9.8\% in the MRR@20 metric on the BBS 40K dataset, and by 6\% in the Hits@10 metric on the WhatsThatBook dataset. This indicates the effectiveness of our method for book search.

    \item  GBS$^P$ performs slightly better than GBS$^S$. This might be due to differences in the generated chapter-level identifiers. The serial decoding in GBS$^S$ maintains the relative order of chapter-level identifiers with respect to the book-level identifier list, whereas parallel decoding does not. If the book-level identifier list is not ideal, parallel decoding can compensate for some of this.
\end{enumerate*}

\subsection{Ablation study}
\label{subsection:ablation-study}
To validate the effectiveness of each component in GBS, we conduct an ablation study on the BBS 40K and WhatsThatBook datasets.
We focus on GBS$^P$, with its retrieval performance shown in Table \ref{tab:ablation}; ablation results based on GBS$^S$ are shown in Table \ref{tab:ablation-s} and show similar trends. Our findings are as follows:
\begin{enumerate*}[label=(\roman*)]
    \item  When not using diversity-enhanced query augmentation (i.e., 2nd row), wherein the model does not learn the retrieval task, there is a significant decline in retrieval performance compared to GBS$^P$ (i.e., 1st row) on both datasets. This highlights the importance of generated pseudo-queries for the model's learning of relevance. For analysis of their quantities, please refer to Section \ref{sec:num-of-pq}.

    \item   When not using coverage-promoting book identifier augmentation, during the learning of each book's content, the book is divided into multiple segments, and pairs of the segment and the book-level identifier are learned.
    This variant (i.e., the 3rd row) exhibits a pronounced performance drop compared to GBS$^P$ on both datasets. This indicates that relying solely on simple book segments and identifier pairs for indexing is insufficient for the model to fully learn book information, thus validating the necessity and effectiveness of coverage-promoting book identifier augmentation.

    \item   When omitting outline-oriented bi-level positional encoding, using only the backbone model's original relative positional encoding, this variant (i.e., the 4th row) shows some performance decline compared to GBS$^P$ on both datasets. This demonstrates that a single layer of positional encoding is inadequate for representing book information, thereby confirming the need for our outline-oriented bi-level positional encoding for book encoding.

    \item   When removing  outline-oriented retentive attention, using the transformer's default standard MHA in the encoder, this variant (i.e., the 5th row)
    shows a slight performance drop compared to GBS$^P$ on both datasets. This indicates that our outline-oriented retentive attention is indeed beneficial for capturing longer input information.
\end{enumerate*}
Table \ref{tab:ablation-s} shows the ablation results based on GBS$^S$.
The trends are the same as those reported for GBS$^P$.

\begin{table}[t]
    \centering
    \setlength{\tabcolsep}{1.2mm}
    \caption{Ablation study of GBS$^P$ on BBS 40K and WhatsThatBook.}
    \label{tab:ablation}
    \begin{tabular}{@{}l@{~}l@{}cccc@{}}
        \toprule
        & \multirow{2}{*}{\textbf{Method}} &  \textbf{BBS 40K} &  \textbf{WhatsThatBook}\\
        \cmidrule(r){3-3}
        \cmidrule(r){4-4}
        & & \textbf{Hits$@10$}  &  \textbf{Hits$@10$}  \\
        \midrule
        1 & GBS$^P$ &56.7 & 70.7\\
        \midrule
        2 & \hspace{1mm} w/o query augmentation & 50.6 & 64.9\\
        3 & \hspace{1mm} w/o identifier augmentation& 45.3 & 60.8\\
        4 & \hspace{1mm} w/o bi-level positional encoding&52.8 & 65.2 \\
        5 & \hspace{1mm} w/o retentive attention& 53.5 & 67.3\\
        \bottomrule
    \end{tabular}
  
\end{table}

\begin{table}[t]
    \centering
    \setlength{\tabcolsep}{1.2mm}
    \caption{Ablation study of GBS$^S$ on BBS 40K and WhatsThatBook.}
    \label{tab:ablation-s}
    \begin{tabular}{@{}l@{~}l@{}cccc@{}}
        \toprule
        & \multirow{2}{*}{\textbf{Method}} &  \textbf{BBS 40K} &  \textbf{WhatsThatBook}\\
        \cmidrule(r){3-3}
        \cmidrule(r){4-4}
        & & \textbf{Hits$@10$}  &  \textbf{Hits$@10$}  \\
        \midrule
        1 & GBS$^S$ &56.3 & 70.4\\
        \midrule
        2 & \hspace{1mm} w/o query augmentation & 50.2 & 64.5\\
        3 & \hspace{1mm} w/o identifier augmentation& 45.1 & 60.6\\
        4 & \hspace{1mm} w/o bi-level positional encoding&52.4 & 64.8 \\
        5 & \hspace{1mm} w/o retentive attention& 53.3 & 67.1\\
        \bottomrule
    \end{tabular}  
\end{table}

\subsection{Analysis on the input length}

\begin{figure}[t]
    \centering
    \includegraphics[width=0.48\textwidth]{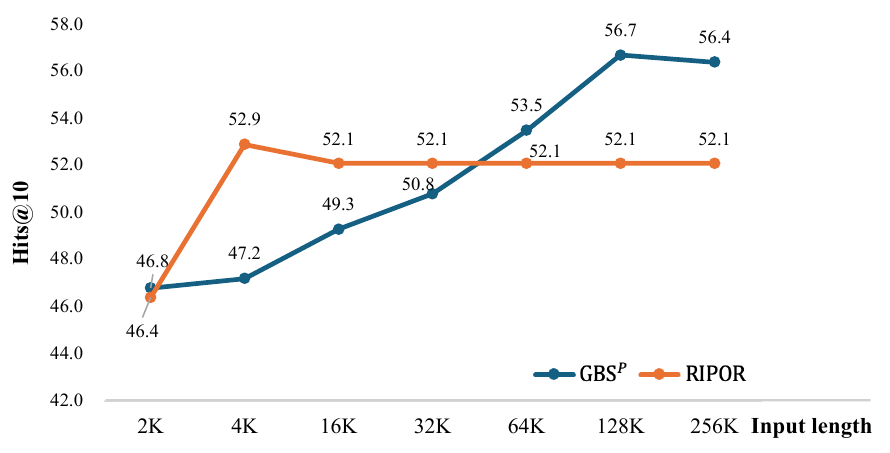}
    \caption{The performance, in terms of Hits@10, of GBS$^P$ and RIPOR with different input lengths on the BBS 40K dataset.}
    \label{fig:input-length}
\end{figure}

To examine how the length of the input text impacts the effectiveness of book search, we vary the input length.
We set input lengths of 2K, 4K, 16K, 32K, 64K, 128K, and 256K on the BBS 40K dataset, and the Hits@10 results for GBS$^P$ and RIPOR are shown in Figure \ref{fig:input-length}. We found that:
\begin{enumerate*}[label=(\roman*)]
    \item   When the input length is between 4K and 32K, RIPOR performs better than our method. This might be because RIPOR's multi-stage optimization and negative sampling strategies allow it to learn important information from the leading parts of the book. Additionally, RIPOR's performance starts to decline once the input length exceeds 4K, and it stabilizes thereafter, indicating that RIPOR has limited capacity for handling extremely long texts like books.

    \item  When the input length is between 64K and 128K, our method outperforms RIPOR, with performance improvements increasing as the input length grows. This suggests that more input content helps the model learn more comprehensive information about the book. We also observed weaker retrieval performance at input lengths of 2K and 4K, which further confirms the complexity of book information and the need for specially designed learning approaches.

    \item  When the input length exceeds 128K, the retrieval performance of GBS$^P$ slightly decreases. This may be because the semantics of the tail parts of some books are included in the leading parts, leading to redundancy and no additional gain.
\end{enumerate*}

\subsection{Impact of the number of pseudo-queries}\label{sec:num-of-pq}
\begin{figure}[t]
    \centering
    \includegraphics[width=0.48\textwidth]{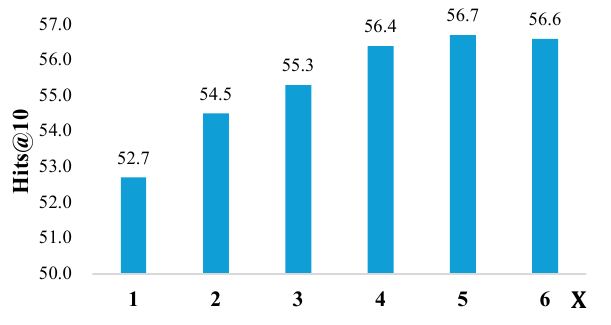}
    \caption{The performance, in terms of Hits@10, of GBS$^P$ with different numbers of diversity-enhanced pseudo-queries, i.e., $X$, on the BBS 40K dataset.}
    \label{fig:num-of-pq}
\end{figure}

Our proposed diversity-enhanced query augmentation strategy generates $X$ pseudo-queries of two types, and the number of these queries is an important factor influencing retrieval performance. Therefore, we analyze the impact of varying $X$.
Specifically, on the BBS 40K dataset, we set $X$ to integer values from 1 to 6, and the Hit@10 performance of GBS$^P$ is shown in Figure \ref{fig:num-of-pq}. We observe the following:
\begin{enumerate*}[label=(\roman*)]
    \item When $X$ is 5 or less, increasing $X$ leads to a larger improvement in retrieval performance. This indicates that pseudo-queries for books help the model better learn relevance and enhance the connections between queries, book information, and identifiers.

    \item When $X$ is set to 6, there is almost no additional gain in retrieval performance. This may be due to model capacity limitations or the fact that $X = 5$ already provides sufficient information for the model to effectively learn book details.
\end{enumerate*}

\subsection{Case study}
To provide a more detailed analysis of the performance of our proposed GBS, we conduct a case study. Specifically, we sample a single-chapter answerable query and one multiple-chapter answerable query from the test set of the BBS 40K dataset and analyze the identifier lists generated by GBS$^P$ and the best baseline, RIPOR.
As shown in Table~\ref{tab:case-study} in Appendix~\ref{sec:appendix-case-study}:
\begin{enumerate*}[label=(\roman*)]
    \item For the single-chapter answerable query, our GBS$^P$ successfully ranks the relevant identifiers at both the book-level and chapter-level in the top position, while RIPOR only ranks them second. This indicates that our method is more effective for handling this type of query.

    \item For the multiple-chapter answerable query, GBS$^P$ also performs well, whereas RIPOR fails to predict the correct identifier in the top 50 identifiers generated. This highlights that book search is more difficult than general web search, and further validates that our method is more adept at addressing book retrieval tasks.
\end{enumerate*}

%% file: sections/related_work.tex
\section{Related Work}

\heading{Book search}
It involves locating and retrieving books based on user queries. Unlike general web search, which handles diverse documents, book search must navigate complex structures within books, including metadata, outlines, and main text, to match user needs accurately \cite{khusro2014issues}.
Traditional methods often rely on term-based matching \cite{fauzi2017arabic, xie2020sembrs} and indexing to connect queries with book metadata. More advanced techniques use natural language processing (NLP) to understand and interpret both queries and book content, improving relevance \cite{bleiweiss2017hierarchical, risch2018book}.

\heading{Generative retrieval}
GR is a new search paradigm, which integrates the entire corpus into a consolidated model, enabling it to generate relevant docids directly from queries \cite{modelBased,DSI}. To achieve this, it involves two core operations \cite{DSI}: indexing, which learns the relationship between document and their docids, and retrieval, which maps queries to relevant docids. GR has gained increasing attention for its strong performance in various retrieval tasks \cite{DSI,zeng2023scalable,genre,tang-2023-recent,tang2023semanticenhancedsedsi,zeng2024planning}. In this work, we attempt to apply GR to book search, which is a challenging and unexplored task due to the unique characteristics of books.

\heading{Long-text modelling}
Long text modeling \cite{dong2023survey} is essential for processing extensive documents like academic articles and reports, which present challenges due to their length and complexity. Traditional RNNs \cite{yang2016hierarchical}, including LSTMs \cite{cohan2018discourse} and GRUs \cite{yang2016hierarchical}, struggle with capturing long-range dependencies, while transformer-based models \cite{vaswani2017attention}, especially pretrained language models (PLMs), have shown promise but face issues with fixed input lengths and high computational costs. Adapting these models to handle long texts involves addressing preprocessing to fit the context length and designing efficient architectures to manage long-term dependencies and hierarchical structures effectively.
Some research improves long text modeling by enhancing positional encoding to capture intrinsic relationships \cite{he2024two,ruoss2023randomized,zhu2023pose}. Other work focuses on enhancing backbone models, such as introducing additional memory module in transformers to retain crucial long-range information \cite{munkhdalai2024leave}. 

%% file: sections/conclusion.tex
\section{Conclusion}
We have introduced and evaluated GBS, a generative retrieval framework designed specifically for book search. Our approach tackles the unique challenges of books by incorporating data augmentation strategies and outline-oriented encoding techniques. 
Experiments on both the industry Baidu dataset and public dataset, show that GBS significantly outperforms existing state-of-the art methods. This confirms the effectiveness of our method in enhancing book search.

However, there are some limitations that could be improved in the future:
\begin{enumerate*}[label=(\roman*)]
    \item Due to the length of books, we construct multiple data pairs to learn book information effectively, which results in high training costs. In the future, we will explore ways to balance performance and learning costs.

    \item Model capacity is also a factor affecting performance. In the future, we will explore the effects on larger capacity backbones.
\end{enumerate*}

\section*{Acknowledgements}

This work was funded by the National Natural Science Foundation of China (NSFC) under Grants No. 62472408,  the Strategic Priority Research Program of the CAS under Grants No. XDB0680102, the National Key Research and Development Program of China under Grants No. 2023YFA1011602, the Lenovo-CAS Joint Lab Youth Scientist Project, and the project under Grants No. JCKY2022130C039.
This research also was (partially) funded by the Hybrid Intelligence Center, a 10-year program funded by the Dutch Ministry of Education, Culture and Science through the Netherlands Organisation for Scientific Research, \url{https://hybrid-intelligence-centre.nl}, project nr.\ 024.004.022, project LESSEN with project number NWA.1389.20.183 of the research program NWA ORC 2020/21, which is (partly) financed by the Dutch Research Council (NWO), project ROBUST with project number KICH3.LTP.20.006, which is (partly) financed by the Dutch Research Council (NWO), DPG Media, RTL, and the Dutch Ministry of Economic Affairs and Climate Policy (EZK) under the program LTP KIC 2020-2023, and the FINDHR (Fairness and Intersectional Non-Discrimination in Human Recommendation) project that received funding from the European Union's Horizon Europe research and innovation program under grant agreement No 101070212, All content represents the opinion of the authors, which is not necessarily shared or endorsed by their respective employers and/or sponsors.

%% file: sections/appendix.tex
\clearpage
\appendix

\section*{Appendix}

\begin{table*}[!b]

\centering
\caption{Given the query, GBS$^P$ and RIPOR return the top-3 beam. Correct results are displayed in \textit{italics}.}
\label{tab:case-study}

\begin{tabular}{p{1cm} p{5cm} p{5cm} p{5cm}}
\toprule
    \multicolumn{4}{l}{\multirow{1}{*}{\begin{minipage}{0.95\linewidth}\textbf{Single-chapter answerable query}: Who is the author of the book ``The Heart of a Boy''?\end{minipage}}}   \\
    \midrule
   \textbf{Method} &   \multicolumn{2}{c}{\textbf{GBS$^P$}} &  \multicolumn{1}{c}{\textbf{RIPOR}}\\
    \cmidrule(r){1-1}
    \cmidrule(r){2-3}
    \cmidrule{4-4}
    \textbf{Rank} & \textbf{\hspace{9mm}Book-level identifier} & \textbf{\hspace{9mm}Chapter-level identifier}& \textbf{\hspace{14mm}Book identifier}  \\
    \midrule
    
    \multicolumn{1}{c}{1}& \textit{The Heart of a Boy\#Edmondo De Amicis\#Laird \& Lee} & \textit{The Heart of a Boy\#Edmondo De Amicis\#Laird \& Lee\#October\#068834} &17-3-8-11-3-24-6-3-12-76-37-37-43-87-33-68-44-174-38-96-221-56-43-78-43-83-7-8-238-1-44-123\\
    
    \multicolumn{1}{c}{2}&Pig Heart Boy\#Malorie Blackman\#Penguin Random House Children's UK
    &\textit{The Heart of a Boy\#Edmondo De Amicis\#Laird \& Lee\#March\#068815}&
    \textit{17-3-8-33-75-32-123-65-168-38-63-183-211-48-95-63-58-168-43-67-83-65-128-46-37-98-68-43-16-43-76-32}\\
    
    \multicolumn{1}{c}{3}&The Boy with a Broken Heart\#Durjoy Datta \#Penguin Metro Reads&
    \textit{The Heart of a Boy\#Edmondo De Amicis\#Laird \& Lee\#November\#068753}&
    17-3-8-11-67-127-39-105-18-35-15-207-48-53-8-178-157-47-36-85-43-17-43-87-9-4-178-164-105-36-41-38\\  

    \midrule
    \multicolumn{4}{l}{\multirow{1}{*}{\begin{minipage}{0.95\linewidth}\textbf{Multiple-chapter answerable query}: Introducing Enrico\end{minipage}}}   \\
    \midrule

    \multicolumn{1}{c}{1}&\textit{The Heart of a Boy\#Edmondo De Amicis\#Laird \& Lee}&
    \textit{The Heart of a Boy\#Edmondo De Amicis\#Laird \& Lee\#October\#068834}&
    17-3-8-11-67-127-39-105-18-35-15-207-48-53-8-178-157-47-36-85-43-17-43-87-9-4-178-164-105-36-41-38
    \\
    
    \multicolumn{1}{c}{2}&An Introduction to the Basics of Reliability and Risk Analysis\#Enrico Zio\#World Scientific Pub Co Inc&
   \textit{ The Heart of a Boy\#Edmondo De Amicis\#Laird \& Lee\#November\#068883}&
    17-3-8-11-3-24-6-3-12-76-37-37-43-87-33-68-44-174-38-96-221-56-43-78-43-83-7-8-238-1-44-123
    \\
    
    \multicolumn{1}{c}{3}&Enrico Baj: The Artist's Home\#Michael Reynolds\#Skira Rizzoli&
    \textit{The Heart of a Boy\#Edmondo De Amicis\#Laird \& Lee\#December\#068659}&
    17-3-8-11-53-62-58-107-38-95-157-23-52-21-230-68-54-89-167-208-32-57-14-58-3-9-54-16-37-48-61-97
    \\

\bottomrule
\end{tabular}
\end{table*}

\section{Baseline Details}\label{appendix-baselines}
The baseline methods are described as follows:

\textit{Sparse retrieval baselines}:
\begin{enumerate*}[label=(\roman*)]
    \item BM25 \cite{bm25} is a commonly used effective term-based method. We use the Anserini toolkit \cite{Anserini} to implement it. We split the book into multiple segments to index.
    \item DocT5Query \cite{doct5query} expands a document with pseudo-queries predicted by a fine-tuned T5 \cite{raffel2020exploringt5} conditioned on the original document. And then we perform the BM25 retrieval. Here, we take the leading 10 chapters as the input to generate pseudo-queries, since this baseline is difficult to encode the whole text of a book.
 \end{enumerate*}

\textit{Dense retrieval baselines}:
\begin{enumerate*}[label=(\roman*)]
    \item RepBERT \cite{zhan2020repbert} is a dual-encoder model with brute force searching;

    \item DPR \cite{karpukhin2020dense} is a classic BERT-based dual-encoder model using dense embeddings for the input texts.
\end{enumerate*}

\textit{GR baselines}:
\begin{enumerate*}[label=(\roman*)]
    \item DSI \cite{DSI} is the first GR work, which uses semantically structured numbers as docids via a k-means clustering algorithm. 
    
    \item GENRE \cite{genre} uses book titles as docids. It learns the document-docid pairs. 
    
    \item SEAL \cite{seal} uses n-grams as identifiers, and generates identifiers based on FM-index. BART-large is used as the backbone.
    
    \item DSI-QG \cite{zhuang2022bridgingdsiqg} generates pseudo-queries conditioned on the book contents using docT5query~\cite{doct5query} and pairs them with identifiers for training. It uses unique integer strings as identifiers.
    
    \item NCI \cite{NCI} employs semantically structured numbers as identifiers. It trains the model using pairs of pseudo-queries and identifiers, and designs a prefix-aware decoder.
  
    \item Corpusbrain \cite{chen2022corpusbrain} employs unique book titles as identifiers for Wikipedia during pre-training. 

    \item Ultron \cite{zhou2022ultron} employs the product quantization code as identifiers. It starts with pre-training using book piece-docid pairs, followed by supervised fine-tuning with annotated queries and generated pseudo-queries on downstream tasks.

    \item GenRet \cite{sun-2023-learning-arxiv} uses an autoencoder to generate identifiers for books, which compress book contents into identifiers and to reconstruct docids back into book contents. It learns jointly with the retrieval task.
    
    \item NOVO \cite{wang2023novo} selects important words from the book as identifiers. The model is trained through supervised learning with annotated information.

    \item ASI \cite{yang2023auto} introduces an additional linear layer to assist in document generation of docids, and introduces negative sample augmentation.

    \item RIPOR \cite{zeng2023scalable} uses a multi-stage optimization strategy and negative mining technique to train the GR model.

\end{enumerate*}

\section{Case Study}\label{sec:appendix-case-study}
Table \ref{tab:case-study} shows the generated top-3 identifiers produced by GBS$^P$ and RIPOR, given two types of queries.